\documentclass[iop,apj,tighten,numberedappendix,twocolappendix]{emulateapj}

\usepackage{apjfonts}

\newcommand{\hb}{H$\beta$}
\newcommand{\heii}{He {\sc ii}}
\newcommand{\feii}{Fe {\sc ii}}
\newcommand{\fvar}{$F_{\rm var}$}
\newcommand{\kms}{$\rm km~s^{-1}$}
\newcommand{\mbh}{$M_\bullet$}
\newcommand{\mcg}{MCG--6-30-15}
\newcommand{\msun}{$M_{\odot}$}
\newcommand{\oiii}{[O~{\sc iii}]}
\newcommand{\rl}{$R_{\rm BLR}$--$L$}
\newcommand{\rmax}{$r_{\rm max}$}

\slugcomment{Accepted for publication in \textit{The Astrophysical Journal}}

\shorttitle{Black Hole Mass in \mcg}
\shortauthors{Hu et al.}

\begin{document}

\title{
Improving the Flux Calibration in Reverberation Mapping by Spectral Fitting:
Application to the Seyfert Galaxy \mcg
}

\author{Chen Hu\altaffilmark{1},
Jian-Min Wang\altaffilmark{1,2},
Luis C. Ho\altaffilmark{3,4},
Jin-Ming Bai\altaffilmark{5},
Yan-Rong Li\altaffilmark{1},
Pu Du\altaffilmark{1},
Kai-Xing Lu\altaffilmark{6,1}
}

\altaffiltext{1}{Key Laboratory for Particle Astrophysics, Institute of High
Energy Physics, Chinese Academy of Sciences, 19B Yuquan Road, Beijing 100049,
China; huc@ihep.ac.cn}

\altaffiltext{2}{National Astronomical Observatories of China, Chinese Academy
of Sciences, 20A Datun Road, Beijing 100020, China}

\altaffiltext{3}{Kavli Institute for Astronomy and Astrophysics, Peking
University, Beijing 100871, China}

\altaffiltext{4}{Department of Astronomy, School of Physics, Peking
University, Beijing 100871, China}

\altaffiltext{5}{Yunnan Observatories, Chinese Academy of Sciences, Kunming
650011, China}

\altaffiltext{6}{Department of Astronomy, Beijing Normal University, Beijing
100875, China}

\begin{abstract}
  We propose a method for the flux calibration of reverberation mapping
  spectra based on accurate measurement of \oiii\ $\lambda 5007$ emission by
  spectral fitting. The method can achieve better accuracy than the
  traditional method of \citet{vangroningen92}, allowing reverberation mapping
  measurements for object with variability amplitudes as low as $\sim$ 5\%. As
  a demonstration, we reanalyze the data of the Seyfert 1 galaxy \mcg\ taken
  from the 2008 campaign of the Lick AGN Monitoring Project, which previously
  failed to obtain a time lag for this weakly variable object owing to a
  relatively large flux calibration uncertainty. We detect a statistically
  significant rest-frame time lag of $6.38_{-2.69}^{+3.07}$ days between the
  \hb\ and $V$-band light curves. Combining this lag with FWHM(\hb) =
  $1933\pm81$ \kms\ and a virial coefficient of $f$ = 0.7, we derive a virial
  black hole mass of $3.26_{-1.40}^{+1.59}\times10^6$ \msun, which agrees well
  with previous estimates by other methods.
\end{abstract}


\section{Introduction}

Reverberation mapping \citep{blandford82} is the most widely applied method
for measuring black hole masses in active galactic nuclei (AGNs) (see
\citealt{peterson14} for a review). It employs spectroscopic monitoring to
derive the mass from the dynamics of the broad-line region (BLR) clouds that
orbit in the gravitational potential of the black hole. The emission-line
width provides an estimate of the cloud velocity, while the time lag between
the continuum and emission line (usually \hb) variability, caused by the
light-travel time between the central continuum and the clouds, yields the BLR
size. Thus, accurate flux calibration is essential for lag measurement in
reverberation mapping. The most widely used method is the spectral scaling
algorithm of \citet[hereinafter vGW92]{vangroningen92}, based on the
normalization of the \oiii\ $\lambda$5007 emission-line intensity, which is
assumed to be constant over the timescale of monitoring observations
\citep{peterson13}. The uncertainty in flux calibration achieved by this
method can be as large as $\sim$ 3\% \citep{barth15}. This level of
uncertainty may lead to erroneous interpretation of AGN light curves for
objects with weak variability, as shown by \citet{barth16} for Mrk 142, whose
variabilities in the $B$ and $V$ bands are $\lesssim$ 3\% \citet{walsh09}. New
methods for flux calibration in reverberation mapping observations would be
valuable.

The spectral fitting method has been used for emission-line measurements in
several reverberation mapping studies
\citep[e.g.,][]{park12,barth13,barth15,hu15}, and it has proved to be an
improvement in these cases to the traditional integration method. By measuring
the flux of the \oiii\ emission line, spectral fitting also naturally can be
used for flux calibration, with the potential to achieve higher accuracy than
the ``standard'' method of vGW92. However, spectral fitting has not been
adopted yet for flux calibration in previous reverberation studies;
\citet{park12} and \citet{barth13,barth15} still used the method of vGW92 for
flux calibration, while \citet{hu15} used a flux calibration strategy that
does not rely on the \oiii\ emission line but on simultaneous observations of
a comparison star. It is worth revisiting reverberation mapping observations
that previously had poor flux calibration, to test whether spectral fitting
can offer any improvement.

\mcg\ is a well-studied nearby ($z$ = 0.008) Seyfert 1 galaxy
\citep[e.g.,][and references
therein]{reynolds97,mchardy05,marinucci14,lira15}, famous for its broad iron
K$\alpha$ line \citep[e.g.,][]{tanaka95,fabian02,miniutti07,marinucci14}. The
mass of the black hole in \mcg\ has been estimated by many methods
\citep[][and references therein]{mchardy05}, all but reverberation mapping.
\mcg\ was spectroscopically monitored in the optical by the Lick AGN
Monitoring Project (LAMP) in 2008 \citep{bentz09}. This object is very
difficult to observe from Lick Observatory due to its southern declination;
the median air mass was 3.2 throughout the campaign \citep{bentz09}. In
addition, the variability of this object during the entire observing period
was rather weak ($\lesssim$ 4\% in the $B$ and $V$ bands; \citealt{walsh09}),
putting greater demands on very accurate spectroscopic flux calibration.
\citet{bentz09} adopted the method of vGW92, which apparently is not good
enough for this object; no BLR time lag was detected because of its noisy \hb\
light curve.

Here, we revisit the LAMP 2008 data of \mcg, show that both the flux
calibration and emission-line measurement can be improved by spectral fitting,
successfully detect the time lag of the \hb\ emission line relative to the AGN
continuum, and derive the mass of the central black hole of this important
source. In Appendix \ref{sec-host}, we also analyze archival {\it Hubble Space
Telescope} ({\it HST}) Wide Field Camera 3 (WFC3) images of \mcg, determine
its bulge classification, and derive the starlight contribution to the
spectroscopic flux.

\section{Flux Calibration}

The details of the observations and data reductions of LAMP 2008 were
presented by \citet{walsh09} and \citet{bentz09}. The photometric monitoring
of \mcg\ included 48 epochs of $B$-band observations with a median cadence of
1.08 days and 55 epochs of $V$-band data with a median cadence of 1.04 days.
The spectroscopy covered 42 epochs with a median cadence of 1.00 days. LAMP
2008 publicly released two sets of spectra: one flux calibrated in the usual
manner using standard stars, and another by scaling to a common flux for the
narrow \oiii\ $\lambda$5007 line, following the procedure of vGW92. Because of
variable sky transparency and slit losses due to seeing and mis-centering,
standard flux calibration is usually not accurate enough for reverberation
mapping. Further refinement in the calibration can be accomplished by noting
that the flux of \oiii\ should be constant over the timescales of the
experiment \citep{peterson13}. The \hb\ light curves in Bentz et al. (2009)
were obtained from the scaled spectra, but the line fluxes were measured by
simple integration of the line after subtraction of a linear, locally defined
continuum.

\begin{figure}
  \centering
  \includegraphics[width=0.45\textwidth]{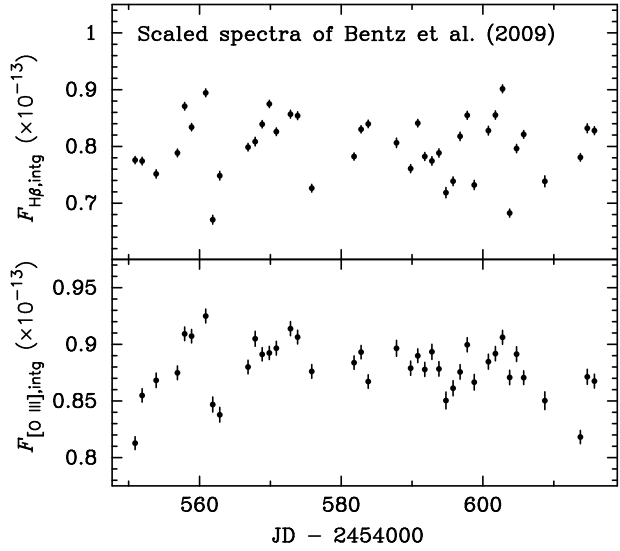}
  \caption{
  Light curves of \hb\ (top) and \oiii\ (bottom) measured from the scaled
  spectra by simple integration, as in \citet{bentz09}, but without plotting
  the systematic error. The scatter in the \oiii\ light curve ($\sim$3\%)
  represents the flux calibration accuracy of the scaled spectra.
  }
  \label{fig-scaled}
\end{figure}

For comparison, we repeat the measurement of the \hb\ light curve of \mcg\ in 
\citet{bentz09}, using the scaled spectra with the same continuum and line
windows. Our results, shown in the top panel of Figure \ref{fig-scaled}, agree
well with those in \citet{bentz09}. Note that the error bars here account 
only for measurement errors, whereas the much larger values in Figure 3 of 
\citet{bentz09} include systematic errors in the light curve. We show below
that this systematic error can be reduced.

\subsection{Flux Calibration Accuracy of the Scaled Spectra}

As the final, calibrated spectra are expected to have the same \oiii\ flux,
the scatter in the \oiii\ flux can be used to estimate the accuracy of the 
flux calibration. The bottom panel of Figure \ref{fig-scaled} shows the \oiii\
light curve measured by integrating the flux above the continuum in the
wavelength range 5030--5065 \AA, using the same continuum windows as for \hb.
The scatter in \oiii\ flux is $\sim$3\%. This accuracy is comparable to that
normally achieved in other reverberation mapping observations, $\sim$2\%
\citep[e.g.,][]{peterson98a,kaspi00}. However, the scatter in \hb\ for this
object is only $\sim$7\%. Moreover, the two light curves are apparently
correlated; their cross-correlation function (CCF) has a peak value \rmax\
$\approx$ 0.7. The scaled spectra are not calibrated accurately enough for
studying the variability of \hb\ flux in this object.

The spectral scaling algorithm of vGW92 does not depend on
the measurement of the narrow-line flux, in order to avoid the uncertainty in
the determination of the continuum, which is significant when the continuum is
defined as a simple straight line set between two wavelength windows.
\citet{hu15} illustrate that spectral fitting enables the continuum to be
properly defined and line fluxes to be accurately measured. Thus, we
recalibrate the reduced spectra by scaling to the \oiii\ flux measured from an
initial spectral fitting. Then, we perform a second, more refined fitting on
the recalibrated spectra to measure the light curves. Before the initial
fitting, we correct the reduced spectra for Galactic extinction using the
$R_V$-dependent law of \citet{cardelli89} and \citet{odonnell94}. We assume
$R_V$ = 3.1 and adopt the $V$-band extinction of 0.165 mag from the NASA/IPAC
Extragalactic Database. All the spectra and light curves hereafter are plotted
after correcting for Galactic extinction.

\subsection{Recalibration of the Reduced Spectra}
\label{sec-recal}

Due to signal-to-noise ratio (S/N) considerations, \citet{hu15} fitted each
individual spectrum by fixing the values of some parameters to those obtained
from the best fit to the mean spectrum. The fixed parameters include the
spectral index of the power-law continuum, the velocity width and shift of the 
broad \heii\ line, and the strength of narrow emission lines. As the mean
spectrum will not be calculated before the recalibration, here we omit the 
broad \heii\ and weak narrow lines from the preliminary fit, only including a
single power-law continuum, the host galaxy, \feii\ emission, broad and narrow
\hb, and \oiii\ $\lambda\lambda$4959, 5007. Accordingly, the preliminary fit
is performed in relatively narrow wavelength windows, covering rest-frame
4430--4600 and 4750--5550 \AA\ (see hatched regions in Fig. \ref{fig-spec}).
The narrow \hb\ line is constrained to have the same width and shift as \oiii,
while all the other parameters are set free. As shown in Section \ref{sec-lc},
this preliminary fitting yields \oiii\ fluxes accurate enough for the purpose
of flux calibration.

Then, for each reduced spectrum, a scaling factor is calculated by dividing a
fiducial flux of $1.02\times10^{-13}$ erg s$^{-1}$ cm$^{-2}$ (obtained from
the \oiii\ $\lambda$5007 flux of $0.856\times10^{-13}$ erg s$^{-1}$ cm$^{-2}$
listed in Table 3 of \citealt{bentz09} after correcting for Galactic
extinction) by our measured \oiii\ flux. By simple scaling of this factor, we
obtain recalibrated spectra for the following light curve measurements and
analysis.

\section{Light Curve Measurements and Analysis}
\label{sec-lc}

\begin{figure}
  \centering
  \includegraphics[angle=-90,width=0.475\textwidth]{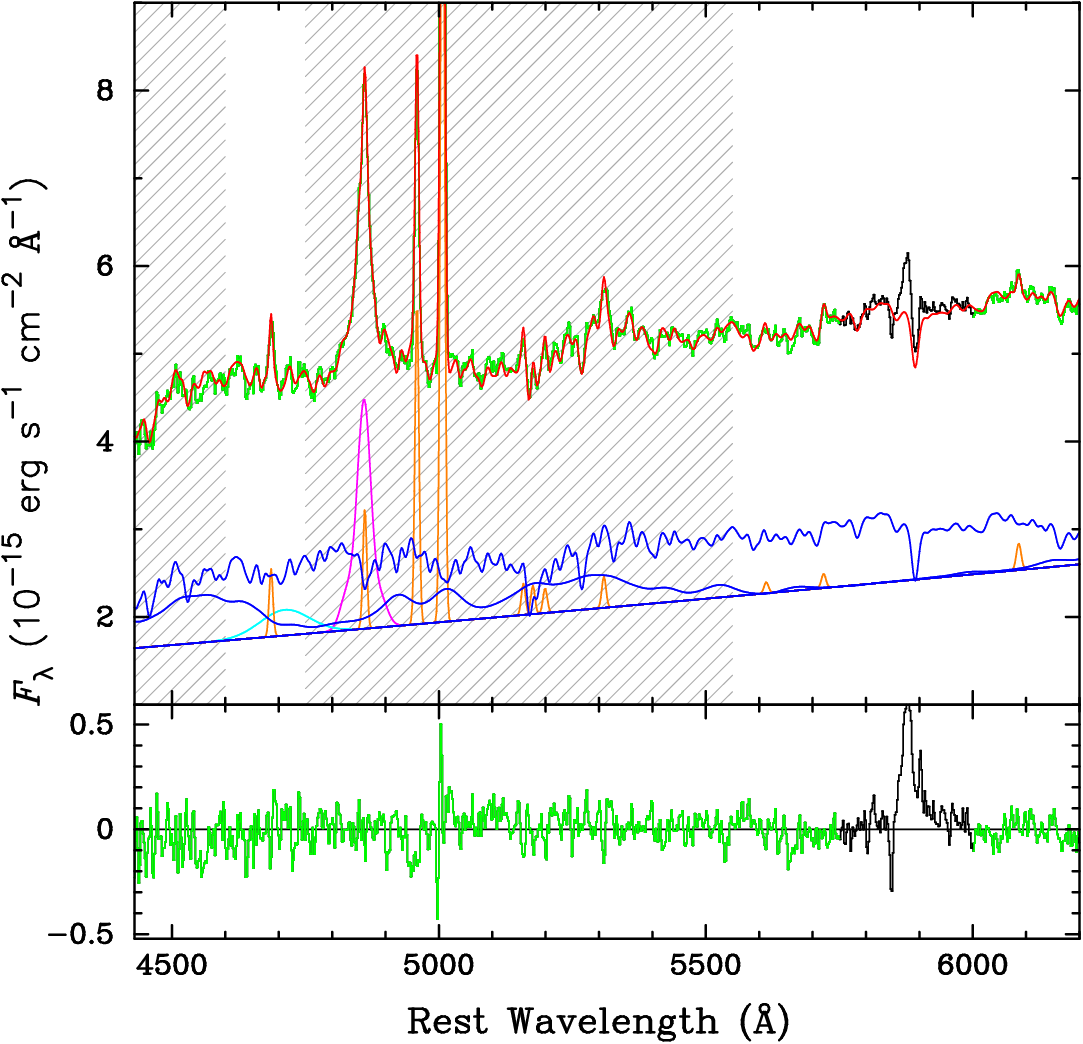}
  \caption{
  Sample fit of a recalibrated spectrum. Pixels included in the fit are
  plotted in green, whereas excluded pixels are in black. The best-fit model
  (red) is composed of the AGN power-law continuum (blue), \feii\ emission
  (blue; template from \citealt{boroson92}), host galaxy (blue; template with
  11 Gyr age and metallicity $Z$ = 0.05 from \citealt{bruzual03}), broad \hb\
  (magenta), broad \heii\ $\lambda$4686 (cyan), and several narrow emission
  lines (orange). The bottom panel shows the residuals. The hatched regions
  show the wavelength windows in the preliminary fitting for the flux
  calibration.
  }
  \label{fig-spec}
\end{figure}

Following \citet{hu15}, we generate light curves from the best-fit values of
the corresponding parameters obtained from the second, more refined spectral
fitting of the recalibrated spectra. Figure \ref{fig-spec} shows an example of
the fit. Compared to the preliminary fitting in Section \ref{sec-recal}, broad
and narrow \heii\ $\lambda$4686 and several narrow coronal lines are added,
and some parameters are fixed to their values obtained from the best fit to
the mean spectrum. Note that the apparent flux variation of the host galaxy
described in Appendix A of \citet{hu15} also exists here, because the size of
the \oiii-emitting region is different from that of the host galaxy. This is
evident from the absorption features in the root-mean-square (rms) spectrum in
Figure 6 of \citet{bentz09}. Thus, as in \citet{hu15}, the fit allows the flux
of the host galaxy to vary. In total, there are 16 free parameters, along with
13 others fixed. We remove from the analysis seven spectra with S/N $<$ 40
(calculated around rest-frame wavelength 5100 \AA) and another four with
reduced $\chi^2$ $>$ 2.4.

The only difference between the procedure here and that of \citet{hu15} is
that we now include narrow \hb\ in the fit, because it can be decomposed well
in the mean spectrum. For each individual-night spectrum, we constrain the
velocity width and shift of narrow \hb\ to be the same as those of \oiii, and
we fix the intensity ratio of narrow \hb\ to \oiii\ $\lambda$5007 to 0.13, as
given by the best-fit mean spectrum. This intensity ratio is consistent with
that measured by \citet{reynolds97} from their nuclear spectrum of \mcg. For
each individual-night spectrum, both the FWHM and the dispersion ($\sigma$) of
broad \hb\ are calculated from the best-fit Gauss-Hermite model; their mean
and standard deviation are used as the measurement of the broad \hb\ width and
its associated uncertainty. After correcting an instrument broadening of 12.5
\AA\ given by \citet{bentz09}, we obtain FWHM = 1933 $\pm$ 81 \kms\ and
$\sigma$ = 1175 $\pm$ 78 \kms. As a consistency check, we also analyzed
archival {\it HST} Space Telescope Imaging Spectrograph (STIS) spectra of
\mcg, in which the broad and narrow \hb\ components are clearly separated (see
Fig. 13 of \citealt{mchardy05}). The FWHM and $\sigma$ of the broad \hb\ from
the STIS spectra are $1935\pm17$ and $1152\pm10$ \kms, respectively, in good
agreement with the measurements obtained above. This demonstrates the
robustness of our spectral decomposition and correction for instrument
broadening.

{\renewcommand{\arraystretch}{1.4}
\begin{deluxetable}{lc}
  \tablewidth{0pt}
  \tablecolumns{2}
  \tablecaption{Measured and Derived Properties of \mcg
  \label{tab-meas}}
  \tablehead{
  \colhead{Parameter} & \colhead{Value}
  }
  \startdata
  FWHM (\hb)                  &  1933 $\pm$ 81 \kms \\
  $\sigma$ (\hb)              &  1175 $\pm$ 78 \kms \\
  \fvar\ (\hb)                &  5.3 $\pm$ 1.1 \% \\
  \rmax\ (\hb\ vs. $V$)       &  0.55 \\
  $\tau$ (\hb\ vs. $V$)       &  $6.38_{-2.69}^{+3.07}$ days \\
  $c \tau {\rm FWHM}^2 / G$   &  $4.66_{-2.00}^{+2.28}\times10^6$ \msun \\
  $c \tau \sigma^2 / G$       &  $1.72_{-0.76}^{+0.86}\times10^6$ \msun \\
  \mbh\ (FWHM, $f=0.7$)       &  $3.26_{-1.40}^{+1.59}\times10^6$ \msun \\
  \mbh\ ($\sigma$, $f=3.2$)   &  $5.51_{-2.43}^{+2.75}\times10^6$ \msun
  \enddata
\end{deluxetable}
}

\begin{figure}
  \centering
  \includegraphics[width=0.45\textwidth]{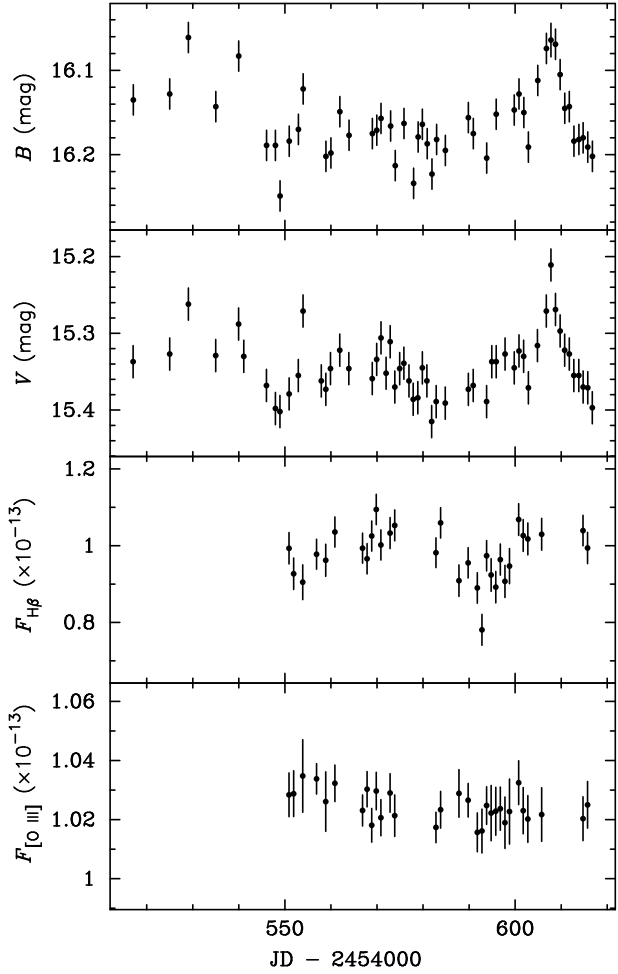}
  \caption{
  Light curves in the $B$ and $V$ bands (top and upper-middle; from
  \citealt{walsh09}), broad \hb\ (lower-middle), and \oiii\ (bottom). Note the
  significant improvement in our flux calibration, shown by the much smaller
  scatter in the \oiii\ flux here, as compared to Figure \ref{fig-scaled}.
  }
  \label{fig-lc}
\end{figure}

The scatter of the measured \oiii\ flux (bottom panel of Fig. \ref{fig-lc}) is
only $\sim$0.5\%, which is less than the fitting error; this indicates that
our recalibration successfully achieved its goal, and that there is no need to
scale further. The light curve of \hb\ is shown in the lower-middle panel. The
error bars of the \hb\ flux include both the fitting error and a systematic
error estimated from the scatter in the measured fluxes of successive nights.
The variability amplitude \fvar\ \citep{rodriguez97,edelson02} of \hb\ is 5.3
$\pm$ 1.1 \%. The top and upper-middle panels show the $B$-band and $V$-band
photometric light curves of \citet{walsh09}. As in \citet{bentz09}, we use the
photometric light curves as the continuum light curve in the time-series
analysis; the light curve of the power-law flux density derived from the
spectral fit has large scatter.

\begin{figure}
  \centering
  \includegraphics[width=0.4\textwidth]{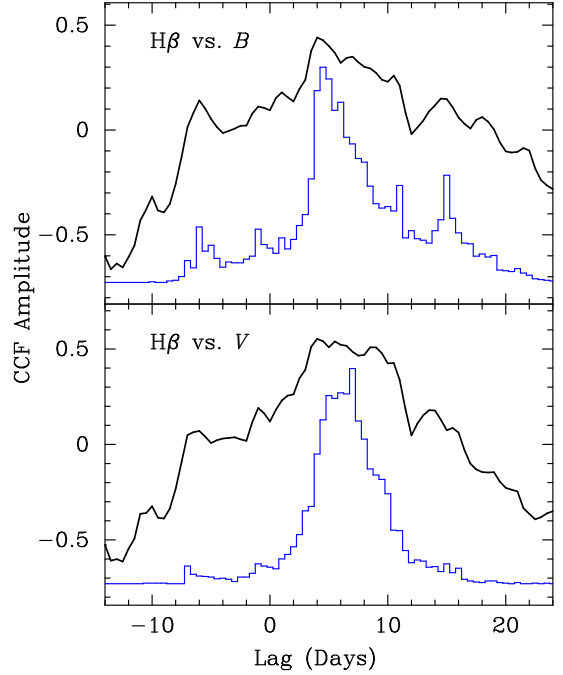}
  \caption{
  Cross-correlation functions (black solid lines) for \hb\ with respect to the
  $B$ (top) and $V$ (bottom) band. The blue histogram in each panel is the
  cross-correlation centroid distribution, which represents the error of the
  time lag.
  }
  \label{fig-ccf}
\end{figure}

We calculate the CCF using the interpolation cross-correlation function method
\citep{gaskell86,gaskell87,white94}, adopting the centroid above 80\% of the
peak value (\rmax) as the time lag \citep{koratkar91,peterson04}. Following
\citet{maoz89} and \citet{peterson98b}, the CCF calculation is repeated for
5000 Monte Carlo realizations, in each of which a random subset of data points
is selected and the fluxes are modified by random Gaussian deviates based on
their errors. The 15.87\% and 84.13\% quantile of the yielded
cross-correlation centroid distribution (CCCD) are used as the lower and upper
uncertainty bounds of the time lag. Figure \ref{fig-ccf} shows the CCFs (black
solid lines) for \hb\ with respect to the $B$ (top) and $V$ (bottom) bands and
the corresponding CCCDs (blue histograms). The rest-frame time lag between
\hb\ and the $V$-band continuum is $\tau = 6.38_{-2.69}^{+3.07}$ days, with
\rmax\ = 0.55. The CCF with respect to the $B$ band gives a consistent time
lag but with large uncertainty ($\tau = 4.45_{-2.49}^{+9.44}$ days), as a
consequence of the large scatter in the $B$-band light curve. Note that the
time lag between \hb\ and the broad-band continuum is expected to be different
from that between \hb\ and the AGN continuum at 5100 \AA, for two reasons.
First, in a thin accretion disk different radii emit continuum radiation
peaked at different wavelengths, as a result of which the lag varies with
wavelength as $\tau \propto \lambda^{4/3}$ (\citealt{collier98}). Second, the
broad-band light curves are contaminated by broad-emission lines. Of the two
bands, $V$ has a pivot wavelength closer to 5100 \AA\ but is more contaminated
by line emission. Both effects have been reported previously. For example,
\citet{fausnaugh16} found a lag of $\sim 0.6$ days between the $B$-band and
$V$-band light curves of NGC 5548, and effect attributable to the
contamination of the continuum bands by broad emission lines; this affects the
inferred broad-band lag by $\sim 0.6-1.2$ days. In the case of \mcg, which has
a smaller black hole mass and lower luminosity, such systematic effects are
expected to be smaller than in NGC 5548 and cannot be distinguished given the
much larger uncertainties in the measured time lags. The following discussion
adopts the time lag with respect to the $V$ band, because the variation in the
$V$ band is stronger than that in the $B$ band and gives a time lag with
smaller uncertainty. (Note that for the other objects in LAMP 2008 the
$B$-band light curve typically has stronger variation and thus was used to
determine the time lag in \citealt{bentz10}.)

Reverberation mapping observations of \hb\ have established an empirical
relation between the BLR radius and luminosity (\rl)
\citep[e.g.,][]{kaspi00,bentz13,du15}. Subtracting the starlight contribution
based on the {\it HST} WFC3 images of \mcg\ (Appendix \ref{sec-host}) yields
an AGN flux density at 5100 \AA\ of $\sim1.1\times10^{-15}$ erg s$^{-1}$
cm$^{-2}$ \AA$^{-1}$. For a luminosity distance\footnote{
Based on the following cosmological parameters: $H_0=72~{\rm km~s^{-1}}$
Mpc$^{-1}$, $\Omega_{m}=0.3$, and $\Omega_\Lambda=0.7$.}
of 33 Mpc, we obtain a spectral luminosity of $\lambda L_{\lambda}$(5100 \AA)
$\approx 7.1\times10^{41}$ erg s$^{-1}$, which, from the \rl\ relation of
\citet{bentz13}, predicts $R_{\rm BLR} \approx 2.4$ lt-days. The Balmer
decrement and color of the power-law continuum suggest that \mcg\ may be
heavily dust reddened. Adopting a reddening of $E(B-V)=0.61$ mag
\citep{reynolds97} yields a much higher intrinsic luminosity of $\lambda
L_{\lambda}$(5100 \AA) $\approx 4.9\times10^{42}$ erg s$^{-1}$ and $R_{\rm
BLR} \approx 6.7$ lt-days. Considering the uncertainties in both the time lag
and reddening, our results show no evidence that \mcg\ deviates from the \rl\
relation. 

\section{Discussion}

\subsection{Mean and rms Spectra}
\label{sec-rms}

In reverberation mapping studies, mean and rms spectra are generated for
measuring the emission-line width. The rms spectrum is preferred for
representing the variable part of the emission line (\citealt{peterson04}; but
see \citealt{barth15} for biases). However, as mentioned in Section
\ref{sec-lc}, the rms spectrum of \mcg\ shows \hb\ absorption instead of
emission because the apparent spectral variability is dominated by variations
in the level of host galaxy contamination. Here, following \citet{barth15}, we
produce two sets of mean and rms spectra, an original set generated in the
standard manner and another generated after subtracting the best-fit AGN
power-law continuum and host galaxy component from each individual spectrum.

Because only a scaling factor was applied to the reduced spectra (see Section
\ref{sec-recal}), our recalibrated spectra still suffer from small nightly
offsets in wavelength shift and spectral resolution. Skipping the corrections
for these offsets does not influence our measurements afterwards because the
widths and shifts of both \hb\ and \oiii\ are free to vary in our fitting; it
conveniently allows us calculate the error of the recalibrated spectrum.
However, the rms spectrum around strong narrow emission lines (e.g., \oiii) is
strongly affected by these nightly offsets. Thus, before generating the mean
and rms spectra, we correct the offsets by applying a linear wavelength shift
and a Gaussian broadening to each individual spectrum according to the \oiii\
velocity shift and width measured in our preliminary fitting used for
recalibration.

\begin{figure}
  \centering
  \includegraphics[width=0.475\textwidth]{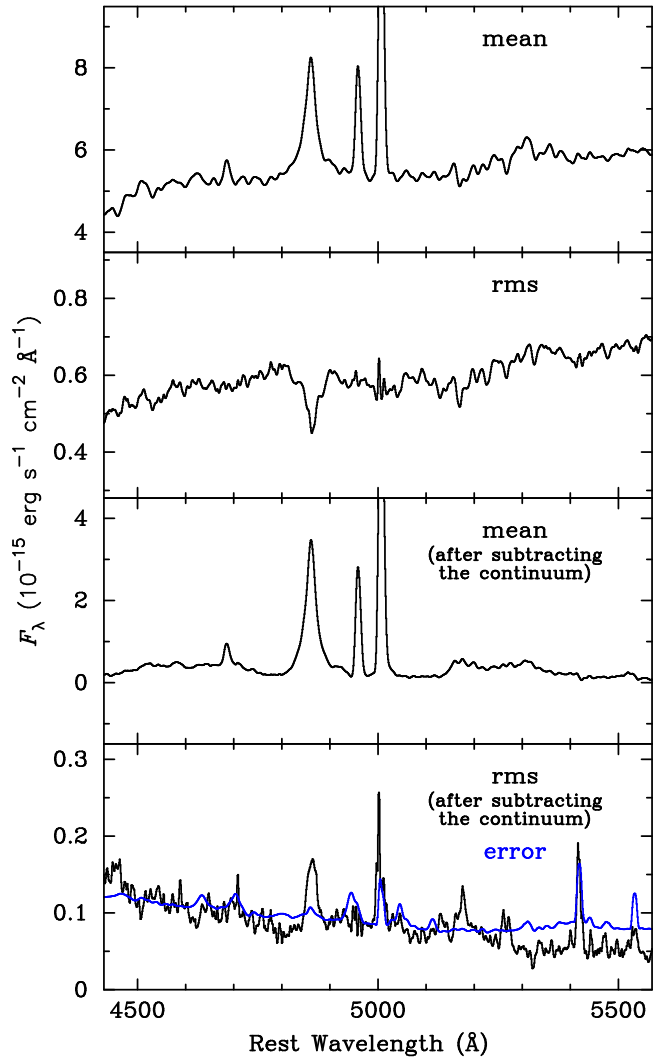}
  \caption{
  Mean and rms spectra generated in the standard manner (top and upper-middle)
  and after subtracting the best-fit AGN power-law continuum and host galaxy
  component from each individual spectrum (lower-middle and bottom). The blue
  curve in the bottom panel is the mean of the error in individual spectrum,
  which would be the rms spectrum if the object is non-varying.
  }
  \label{fig-rms}
\end{figure}

Figure \ref{fig-rms} shows the two sets of mean and rms spectra. The standard
rms spectrum (upper-middle panel) resembles that presented in Figure 6 of
\citet{bentz09}. It shows no \hb\ emission but absorption, because the
variability of the \hb\ line is overwhelmed by the variations in the AGN
power-law continuum and the level of host galaxy contamination. After
subtracting the best-fit continuum components, which includes both the AGN
power-law and the host galaxy from each individual spectrum, the resulting rms
spectrum (black curve in the bottom panel) has much lower intensity than the
original one, and shows emission features of \hb\ and \oiii. The blue curve in
the bottom panel is the mean of the errors of individual-night spectra in each
wavelength bin, which represents the minimal level of rms variations that
could be generated. It is the expected rms spectrum of a non-varying object,
for which the variations in the observed spectra are totally caused by the
observational errors. By comparing the continuum-subtracted rms spectrum and
the error spectrum, the apparent variation of the \oiii\ line can be mostly
attributed to observational error; it also illustrates that our recalibration
works well. On the other hand, the \hb\ line revealed in the
continuum-subtracted rms spectrum is much stronger than the error, which means
that the variability of the \hb\ line is observable and successfully measured
by our spectral fitting. Fitting a single Gaussian with a local continuum to
the \hb\ line in the continuum-subtracted rms spectrum yields a FWHM of 1104
\kms\ (after instrumental broadening correction), which is much narrower than
the mean FWHM of the individual spectrum listed in Table \ref{tab-meas}, a
possible consequence of biases due to the weak variability of the feature (see
\citealt{barth15} for detailed discussions).

\subsection{Black Hole Mass}

Combining the time lag $\tau$ with the emission-line width $\Delta V$ provides
an estimate of the virial mass of the black hole,
\begin{equation}
  M_\bullet = f \frac{c \tau \Delta V^2}{G}~,
\end{equation}
where $c$ is the speed of light, $G$ is the gravitational constant, and $f$ is
a virial factor that absorbs the unknown geometry, kinematics, and orientation
of the BLR. As shown above, the usual way of measuring $\Delta V$ from the rms
spectrum is not suitable for \mcg\ here. The continuum-subtracted rms spectrum
does reveal the variation of the \hb\ emission line but may still be biased
for broad-line width measurement due to the noise in each individual spectrum
(see the simulations in \citealt{barth15}). Thus, we measure FWHM and $\sigma$
from the best-fit model of each individual spectrum, and we compute $\Delta V$
from the mean of the individual measurements over the entire observing
campaign. In practice \citep[e.g.,][]{onken04}, the $f$-factor is calibrated
from the \mbh--$\sigma_{\ast}$ relation of inactive galaxies (see
\citealt{kormendy13} for review), which depends on bulge type, being
systematically lower for pseudobulges than classical bulges \citep{ho14}. With
a bulge-to-total light ratio of 0.06 and a bulge S\'ersic of $n$ = 1.29
(derived from archival {\it HST} WFC3 images; Appendix \ref{sec-host}), \mcg\
appears to host a pseudobulge. Adopting $f$ = 0.7 \citep{ho14} for $\Delta V$
= FWHM, we estimate \mbh\ = $3.26_{-1.40}^{+1.59}\times10^6$ \msun; with
$\Delta V = \sigma$, $f$ = 3.2 and \mbh\ = $5.51_{-2.43}^{+2.75}\times10^6$
\msun. Table \ref{tab-meas} lists all measured and derived quantities for
\mcg.

The black hole mass in \mcg\ previously has been estimated using a variety of
methods, with that from the X-ray power-spectral density technique yielding
smallest uncertainty: $2.9_{-1.6}^{+1.8}\times10^6$ \msun\
(\citealt{mchardy05}). Other estimates, including those from the correlations
between black hole mass and bulge properties and photoionization-based
calculations of the BLR size, all employed optical observations and derived
values between in the range $\sim(3-6)\times10^6$ \msun\ (see the
comprehensive discussion in \citealt{mchardy05}). All these methods are
indirect, based on correlations between certain observables and the black hole
mass. This work is the first successful direct measurement of the black hole
mass in \mcg, based on the dynamics of gas in the gravitational potential of
the black hole itself. Our two measurements, based on FWHM and $\sigma$, are
both in the range of masses given by previous studies, but have smaller
uncertainties.  We prefer the mass based on FWHM
($3.26_{-1.40}^{+1.59}\times10^6$ \msun) because it is in better agreement
with the mass estimate from X-ray variability.

\mcg\ has an \hb\ line width (FWHM $<$ 2000 \kms) that formally qualifies it
as a narrow-line Seyfert 1 galaxy \citep{osterbrock85}, although its \feii\
emission is not as strong and the \oiii/\hb\ ratio is not as low as in most
objects of this class. For a canonical bolometric correction of $L_{\rm bol} =
9.8L_{5100}$ \citep{mclure04}, we obtain an Eddington ratio of $L_{\rm
bol}/L_{\rm Edd}$ = 0.12 using the virial black hole mass based on the \hb\
FWHM and the intrinsic luminosity assuming a reddening of $E(B-V)=0.61$ mag.
Such an Eddington ratio is low among narrow-line Seyfert 1 galaxies (see,
e.g., Figure 1 of \citealt{xu12}), but consistent with the weak \feii\
emission in this object.

\subsection{Remarks on the Calibration and Measurement Methods}

The spectral scaling algorithm of vGW92 is the most widely
used method for flux calibration of reverberation mapping observations. It has
proven to be effective in most cases. However, as mentioned in \citet{barth15}
and shown in Figure \ref{fig-scaled} of this paper, the uncertainty in the
flux calibration achieved can be as large as $\sim$3\%. This is not accurate
enough for studying objects with weak variability, such as \mcg.
\citet{barth15} also find that the vGW92 method does not work
optimally for the objects with low \oiii\ equivalent width. Moreover, \oiii\
$\lambda$5007 is blended with \feii\ $\lambda$5018, which may be strong and
variable (\citealt{hu15}; see their Fig. 3 for examples); under these
circumstances, spectral decomposition is necessary to deblend \oiii. This
paper demonstrates that spectral fitting provides a reliable and
straightforward way to measure robust \oiii\ line strengths, which improves
the flux calibration accuracy to $\sim$0.5\%.

The spectral fitting method has been used more and more to measure light
curves in recent reverberation mapping studies
\citep[e.g.,][]{bian10,barth13,barth15,hu15}. This method is useful for
dealing with blended emission lines such as \feii\ and \heii, and it can also
improve the measurement of \hb\ (see Fig. 6 of \citealt{hu15} for a comparison
with the traditional integration method). For objects with strong host galaxy
contribution, spectral fitting also allows correction for the apparent flux
variation of the host galaxy \citep{hu15}, thereby reducing the contamination
of \hb\ emission by absorption, as in the case of \mcg\ here.

In summary, our success in obtaining a time lag measurement for \mcg\ in this
paper results from improvements in both flux calibration and light curve
measurements. This is accomplished by spectral fitting in two steps: an
initial fit to obtain the \oiii\ flux for calibration, followed by a second
fit to include additional spectral components to extract the light curve of
the broad \hb\ emission line. We have demonstrated for the first time the
power of this technique in flux calibration, and suggest applying it as an
alternative approach in future reverberation studies, especially those of
objects with weak variability.

\acknowledgments
This work made use of data from the Lick AGN Monitoring Project public data
release, and archival data from the NASA/ESA {\it Hubble Space Telescope}. We
appreciate extensive discussions among the members of the IHEP AGN group. We
thank Aaron Barth for reading and commenting on a draft of this paper, and an
anonymous referee for helpful comments and suggestions. This research is
supported by grant 2016YFA0400702 from the Ministry of Science and Technology
of China, by the Strategic Priority Research Program -- The Emergence of
Cosmological Structures of the Chinese Academy of Sciences, grant No.
XDB09000000, by the CAS Key Research Program through KJZDEW-M06, by the NSFC
through NSFC-11173023, -11133006, -11233003, -11473002, -11503026, -11573026,
by a NSFC-CAS joint key grant U1431228, and by a China-Israel project
NSFC-11361140347.

\appendix

\section{Host Galaxy}
\label{sec-host}

The host galaxy of \mcg\ is classified as an S0 galaxy by \citet{malkan98},
based on {\it HST}\ Wide Field and Planetary Camera 2 images that show a
saturated nucleus and a dust lane on one side of major axis. In this Appendix,
we analyze archival {\it HST}/WFC3 images of this galaxy (GO-11662, PI:
Bentz), for two purposes: (1) determining the bulge classification and (2)
measuring the starlight flux contribution to the spectroscopic flux at 5100
\AA.

\begin{figure*}
  \centering
  \includegraphics[width=0.8\textwidth]{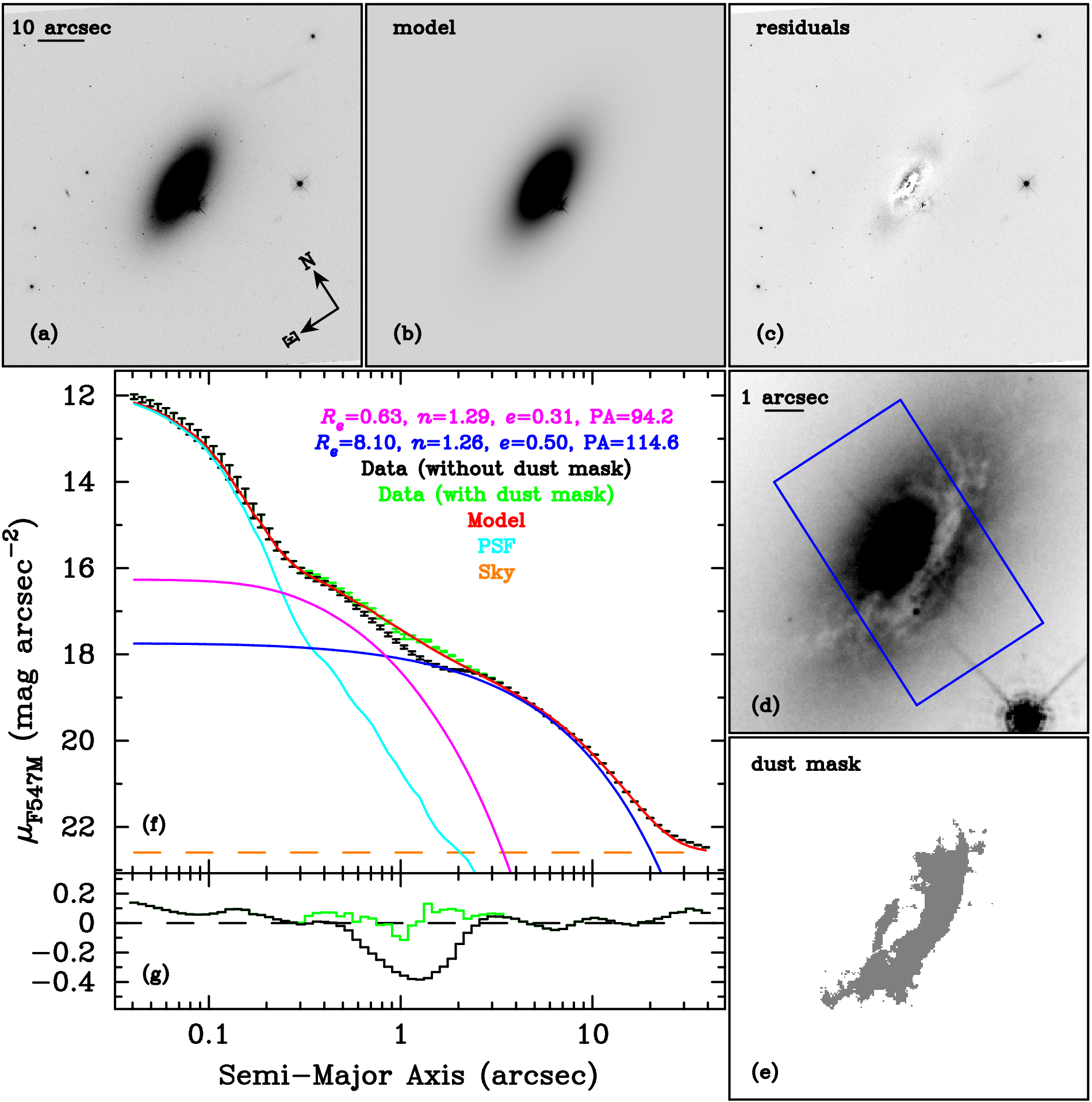}
  \caption{
  (a)--(c) {\it HST} WFC3 F547M image, best-fit model, and residuals for \mcg.
  Panel (d) zooms in around the nucleus to show the dust lane and
  spectroscopic extraction aperture (blue rectangle). (e) The final converged
  dust mask (see text for details). (f) One-dimensional surface brightness
  profile (black points with error bars), the best-fit model (red line), and
  the individual component for the bulge (magenta line) and disk (blue line).
  The PSF is shown as a cyan line. (g) The residuals between the data (with
  and without the dust mask) and the model.
  }
  \label{fig-host}
\end{figure*}

\mcg\ was observed by the WFC3 Ultraviolet-Visible (UVIS) channel with the
F547M filter (see \citealt{bentz13} for the details of the observations). Two
sets of exposures with integration times of 25, 370, and 750 s were taken. We
retrieve the flat-fielded images of the six exposures from the {\it HST}
archives and replaced the saturated pixels in the long-exposure images with
scaled versions of unsaturated pixels from the shallower exposures. Then, we
use the {\tt DrizzlePac} task {\tt AstroDrizzle} (version 1.1.16;
\citealt{gonzaga12}) to clean cosmic rays, correct geometric distortion, and
create the final combined image (Figure \ref{fig-host}a) and error image. We
generate point-spread function (PSF) models using Tiny Tim (version 7.5;
\citealt{krist11}) for each exposure, and combine them in the same way as the
galaxy images.

We use GALFIT (version 3.0.5; \citealt{peng02,peng10}) to perform
two-dimensional surface brightness decomposition. The model includes the
following components: (1) two PSF profiles for the AGN and the nearby bright
star to the south, (2) two \citet{sersic68} profiles for the bulge and disk of
the host galaxy, and (3) a constant background sky. A dust mask is needed for
the central dust lane (Figure \ref{fig-host}d). We generate it iteratively.
We first fit the image without a dust mask and generate an initial mask using
the pixels with negative residual values that exceed 5 times the error. Then
we refit the image with the dust mask and generate a new dust mask based on
the new residuals. After a few iterations, both the dust mask and the best-fit
parameters converge. The final converged dust mask is shown in Figure
\ref{fig-host}(e); its shape resembles that of the fuzzy dust lane.

The final best fit using the converged dust mask has a reduced $\chi^2$ of
0.994. Panels (b) and (c) of Figure \ref{fig-host} show images of the best-fit
model and residuals. Panel (f) shows the one-dimensional surface brightness of
the data (with and without the dust mask), the best-fit model, and each
component of the model, plotted in different colors. Panel (g) shows the
residuals. Table \ref{tab-host} lists the best-fit values of the parameters,
which yield a bulge-to-total ($B/T$) light ratio of 0.06 and a bulge S\'ersic
index $n$ = 1.29. Thus, we classify the bulge of \mcg\ as a pseudobulge based
on its low $B/T$ and $n$ \citep{gadotti09}.

\begin{deluxetable}{lcccccc}
  \tablewidth{0pt}
  \tablecolumns{7}
  \tablecaption{Surface Brightness Decomposition
  \label{tab-host}}
  \tablehead{
  \colhead{Model} & \colhead{$m_{\rm ST}$\tablenotemark{a}} & \colhead{$R_e$}
  & \colhead{$n$} & \colhead{$e$} & \colhead{PA}      & \colhead{Note} \\
  \colhead{}      & \colhead{(mag)}        & \colhead{(\arcsec)}
  & \colhead{}    & \colhead{}    & \colhead{(\degr)} & \colhead{}
  }
  \startdata
  PSF      & 16.53 & \nodata & \nodata & \nodata & \nodata & AGN     \\
  S\'ersic & 16.44 & 0.63    & 1.29    & 0.31    & 94.2    & Bulge   \\
  S\'ersic & 13.42 & 8.10    & 1.26    & 0.50    & 114.6   & Disk    \\
  PSF      & 15.56 & \nodata & \nodata & \nodata & \nodata & Star    \\
  Sky\tablenotemark{b} & 26.22
  & \nodata & \nodata & \nodata & \nodata & \nodata 
  \enddata
  \tablenotetext{a}{The ST magnitude system is based on constant flux per unit
  wavelength. $m_{\rm ST}=-2.5{\rm log}(f_\lambda)-21.10$.}
  \tablenotetext{b}{The units for the sky value is in counts.}
\end{deluxetable}

Using the results of the surface brightness decomposition, we measure the
starlight contribution to the spectroscopic flux, following \citet{bentz13}.
The blue rectangle in Figure \ref{fig-host}(d) shows the aperture
(4\arcsec$\times$7\arcsec; \citealt{bentz09}) in which the LAMP 2008 spectra
were extracted. We subtract the best-fit AGN and sky components from the data
image, and measure the flux within the aperture. Then, we determine the color
correction using the IRAF {\it synphot} package and the galaxy template
adopted in our spectral fitting (Section \ref{sec-lc}). The host galaxy flux
density contribution to the spectra at rest-frame 5100 \AA\ is
$4.3\times10^{-15}$ erg s$^{-1}$ cm$^{-2}$ \AA$^{-1}$, after correcting for
Galactic extinction. For comparison, our spectral decomposition in Section
\ref{sec-lc} yields a somewhat lower average host galaxy flux density of
$3.1\times10^{-15}$ erg s$^{-1}$ cm$^{-2}$ \AA$^{-1}$. The difference may
come from the uncertainty in the spectral decomposition, especially when the
spectral slope of the AGN continuum has strong dust reddening, which is not
tightly constrained \citep{reynolds97}. Also, measuring the flux contribution
from the image has an uncertainty of $\sim$10\% \citep{bentz13}. Finally, we
obtain a starlight-subtracted AGN flux density at 5100 \AA\ of
$1.1\times10^{-15}$ erg s$^{-1}$ cm$^{-2}$ \AA$^{-1}$.

\noindent\begin{minipage}{\textwidth}
\vspace{2em}
\hspace{1em}{\it Note after acceptance}.---\citet{bentz16} report a
determination of \hb\ time lag from a different reverberation campaign. Their
time lag and that in the present paper are consistent.
\end{minipage}

\end{document}